\documentclass[aps,prb]{revtex4}
\usepackage{graphicx}

\begin{document}
\pagestyle{empty}
\title{Near-field radiative heat transfer and van der Waals friction between closely spaced
graphene and amorphous SiO$_2$}

\author{A.I.Volokitin$^{1,2}$\footnote{Corresponding author.
\textit{E-mail address}:alevolokitin@yandex.ru}    and B.N.J.Persson$^1$}
 \affiliation{$^1$Institut f\"ur Festk\"orperforschung,
Forschungszentrum J\"ulich, D-52425, Germany} \affiliation{
$^2$Samara State Technical University, 443100 Samara, Russia}

\begin{abstract}
We study the radiative heat transfer and the van der Waals
friction between graphene and an amorphous SiO$_2$ substrate. We
study the surface phonon-polaritons contribution to  the low-field
mobility  as a function of temperature and  of carrier density. We
find that the electric current saturate at a high electric field,
in agreement with experiment. The saturation current depends
weakly on the temperature, which we attribute to the ``quantum"
friction between the graphene carriers and the substrate optical
phonons. We calculate the frictional drag between two graphene
sheets caused by van der Waals friction, and find that this drag
can induce a high enough voltage which can  be easily measured
experimentally. We find that for nonsuspended graphene the
near-field radiative heat transfer, and the heat transfer due to
direct phononic coupling, are of the same order of magnitude at
low electric field. The phononic contribution to the heat transfer
dominates at high field. For large separation between graphene and
the substrate the heat transfer is dominated by the near-field
radiative heat transfer.
\end{abstract}

\maketitle


\vskip 5mm

\textbf{Introduction.} Graphene, the recently isolated
single-layer carbon sheet, consist of carbon atoms closely packed
in a flat two-dimensional crystal lattice. The unique electronic
and mechanical properties of graphene\cite{Geim2004, Geim2005} is
being actively explored both theoretically and experimentally
because of its importance for fundamental physics, and for
possible technological applications \cite{Geim2007}. In
particular, a great deal of attention has been devoted to the applications
of graphene for electronics and sensoring
\cite{Geim2004,Geim2007}.

Graphene, as for all media, is surrounded  by a fluctuating
electromagnetic field due to the thermal and quantum
fluctuations of the current density.  Outside the bodies
this fluctuating electromagnetic field exists partly in the form
of propagating electromagnetic waves and partly in the form of
evanescent  waves. The theory of the fluctuating electromagnetic
field was developed by Rytov \cite{Rytov53,Rytov67,Rytov89}. A
great variety of phenomena such as Casimir-Lifshitz forces
\cite{Lifshitz54}, near-field radiative heat transfer
\cite{Polder1971}, non-contact friction
\cite{Volokitin99,VolokitinRMP07,VolokitinUFN07,Volokitin2008b},
and the frictional drag in low-dimensional systems
\cite{Volokitin2001b,Volokitin2008a}, can be described using this
theory.

In the present paper we apply the theories of the near-field
radiative heat transfer and the van der Waals friction to study
the electric transport properties, and the heat generation and
dissipation in graphene, mediated by the fluctuating
electromagnetic field. The charge carriers in graphene absorbed
on, or located close to, a substrate experience a friction due to
interaction with substrate. This friction affects the graphene
carrier mobility and  leads to the heat generation. The existing
microscopic theories\cite{Perebeinos2009,Perebeinos2010} involve
several fitting parameters, while our theory is macroscopic. In
this theory  the electromagnetic interaction between graphene and
a substrate is described by the  dielectric functions of the
materials which can be accurately determined from experiment.

The development of more powerful microelectronic devices results in
increasing dissipated power (per unit volume), and developing structures with smaller
heat production and better cooling strategies,
is one of the ``grand challenge'' of modern electronics
\cite{Handbook2007}. The heat generated in microelectronic devices
leads to elevated device operation
temperatures, performance reduction, and ultimately to hardware failures.
The three classical mechanisms of heat transfer are
convection, conduction, and radiation. The former is not important
in existing circuit architecture having no fluid or gas flowing
inside.

Graphene interact very weakly with most substrates mainly via van der
Waals forces. According to theoretical calculations\cite{Persson2010}
the thermal contact conductance due to the direct
phononic coupling for the interface between graphene
and a perfectly smooth (amorphous) SiO$_2$ substrate is $\alpha_{\rm ph} \approx
3\times 10^8$Wm$^{-2}$K$^{-1}$, and according to experiment\cite{Chen2009}
(at room temperature) the thermal contact conductance ranges from 8$\times$10$^7$ to
1.7$\times$10$^8$Wm$^{-2}$K$^{-1}$ (however, these values are probably influenced by the
substrate surface roughness).

At large separation $d\gg \lambda_T=k_BT/\hbar$ the radiative heat
transfer is determined by the Stefan-Boltzman law, according to
which $\alpha=4\sigma T^3$. In this limiting case the heat
transfer between two bodies is determined by the propagating
electromagnetic waves radiated by the bodies, and does not depend
on the separation $d$. At $T=300$K this law predicts the (very
small)  thermal contact conductance, $\alpha \approx 6$
Wm$^{-2}$K$^{-1}$. However, as was first predicted theoretically
\cite{Polder1971}, and recently confirmed experimentally
\cite{Chen2009a,Greffet2009}, at short separation $d\ll
\lambda_T$, the heat transfer may increase by many orders of
magnitude due to the evanescent electromagnetic waves; this is
often referred to as photon tunneling. Particularly strong
enhancement occurs if the surfaces of the bodies can support
localized surface modes such as surface plasmon-polaritons,
surface phonon-polaritons, or adsorbate vibrational modes
\cite{Greffet6,VolokitinRMP07}. In two recent experiments
\cite{Chen2009a,Greffet2009} the thermal contact conductance
between a silica plate and a silica sphere was measured from 30 nm
separation out to few micrometers. Heat transfer across the
plate-sphere gap causes the cantilever to bent very slightly, and
this was measured by optical fiber interferometry.

The theory of the radiative heat transfer developed in Ref.
\cite{Polder1971} is only valid for bodies at rest. A more general
theory of the radiative heat transfer between moving bodies, with
arbitrary relative velocities, was developed in Ref.
\cite{Volokitin2008b}. This theory can be applied to calculate the
radiative heat transfer between carriers (moving with the drift
velocity $v$) in graphene and the substrate. According to theory
\cite{VolokitinRMP07}, for the graphene-SiO$_2$ interface the
maximum of the contribution of the radiative heat transfer to the
thermal contact conductivity is $\alpha_{max}\sim
(k_BT)^2n/\hbar$, where $n$ is the concentration of carriers in
graphene. At room temperature and $n\approx 10^{13}$cm$^{-2}$ this
gives $\alpha_{max}\sim 10^7$Wm$^{-2}$K$^{-1}$. However this
estimation does not take into account that the charge carriers in
graphene are moving relative to the substrate. Relative motion of
charge carriers in graphene  corresponds to an effective increase
in the temperature difference between graphene and substrate, what
leads to an increase in thermal contact conductance.  Thus, for
nonsuspended graphene the thermal contact conductivity due to
near-field radiative heat transfer, and due to direct  phononic
mechanism, can be of the same order. However for suspended
graphene, the main contribution results from radiative heat
transfer since this process exhibit a much weaker distance
dependence than for the phononic contribution.

\begin{figure}
\includegraphics[width=0.45\textwidth]{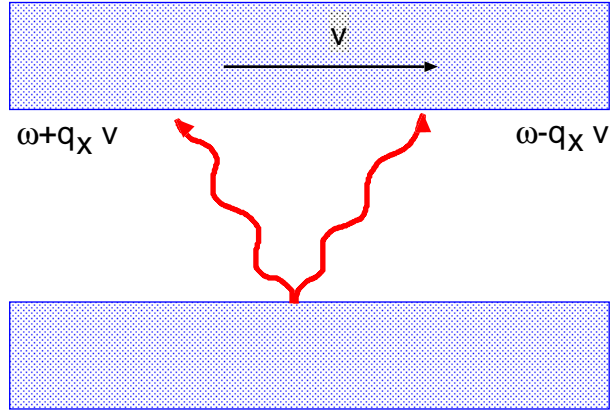}
\caption{\label{Fig1} Two bodies moving relative to each other
will experience van der Waals friction due to Doppler shift of the
electromagnetic waves emitted by them. }
\end{figure}

The radiative heat transfer is closely connected with the van
der Waals friction. This friction is due to the Doppler effect.
Assume that a graphene sheet is separated from the
substrate by a sufficiently wide
insulator gap, which prevents particles from tunneling across it.
If the charge carriers inside graphene move with velocity  $v$
relative to the substrate, a frictional stress will act on them.
This frictional stress is related to an asymmetry of the
reflection amplitude along the direction of motion; see Fig. \ref{Fig1}.
If the substrate emits radiation, then in the rest reference frame of
charge carriers in graphene these waves are Doppler shifted which
will result in different reflection amplitudes. The same is true
for radiation emitted by moving charge carriers in graphene. The
exchange of ``Doppler-shifted-photons'' will result in momentum
transfer between graphene and substrate, which is the origin of
the van der Waals friction.

\textbf{Theory.} Let us consider graphene and a substrate with
flat parallel surfaces at separation $d\ll \lambda_T=c\hbar/k_BT$.
Assume that the free charge carriers in graphene move  with the
velocity $v\ll c$  ($c$ is the light velocity) relative to other
medium. According to Ref.
\cite{Volokitin99,VolokitinRMP07,VolokitinUFN07,Volokitin2008b}
the frictional stress $F_x$ acting on charge carriers in graphene,
and the radiative heat flux $S_z$ across the surface of substrate,
both mediated by a fluctuating electromagnetic field, are determined by

\[
F_x =\frac \hbar {\pi ^3}\int_{0 }^\infty dq_y\int_0^\infty
dq_xq_xe^{-2qd}\Bigg \{ \int_0^\infty d\omega \Bigg(
\frac{\mathrm{Im}R_{d}(\omega)\mathrm{Im}R_{g}(\omega^+) }{\mid
1-e^{-2 q d}R_{d}(\omega)R_{g}(\omega^+)\mid ^2}\times
\]
\[
 [n_d(\omega )-n_g(\omega^+)]+\frac{\mathrm{Im}R_{d}(\omega^+)\mathrm{Im}R_{g}(\omega^) }{\mid
1-e^{-2 q d}R_{d}(\omega^+)R_{g}(\omega)\mid ^2}[n_g(\omega
)-n_d(\omega^+)]\Bigg )+
\]
\begin{equation}
 \int_0^{q_xv}d\omega \frac{\mathrm{Im}R_{d}(\omega)\mathrm{Im}
R_{g}(\omega^-)} {\mid 1-e^{-2qd}R_{d}(\omega)R_{g}(\omega^-)\mid
^2} [n_g(\omega^-)-n_d(\omega)] \Bigg \}, \label{parallel2}
\end{equation}
\[
S_z =\frac \hbar {\pi ^3}\int_{0 }^\infty dq_y\int_0^\infty
dq_xe^{-2qd}\Bigg \{ \int_0^\infty d\omega \Bigg(- \frac{\omega
\mathrm{Im}R_{d}(\omega)\mathrm{Im}R_{g}(\omega^+) }{\mid 1-e^{-2
q d}R_{d}(\omega)R_{g}(\omega^+)\mid ^2}\times
\]
\[
 [n_d(\omega )-n_g(\omega^+)]+\frac{\omega^+\mathrm{Im}R_{d}(\omega^+)\mathrm{Im}R_{g}(\omega^) }{\mid
1-e^{-2 q d}R_{d}(\omega^+)R_{g}(\omega)\mid ^2}[n_g(\omega
)-n_d(\omega^+)]\Bigg )+
\]
\begin{equation}
 \int_0^{q_xv}d\omega \frac{\omega \mathrm{Im}R_{d}(\omega)\mathrm{Im}
R_{g}(\omega^-)} {\mid 1-e^{-2qd}R_{d}(\omega)R_{g}(\omega^-)\mid
^2} [n_g(\omega^-)-n_d(\omega)] \Bigg \}, \label{parallel2}
\end{equation}
where  $n_i(\omega )=[\exp (\hbar \omega /k_BT_i-1]^{-1}$
($i=g,d$), $T_{g(d)}$ is the temperature of graphene (substrate),
 $R_{i}$  is the reflection amplitude for
surface $i$ for $p$ -polarized electromagnetic waves, and
$\omega^{\pm}=\omega \pm q_xv$.  The reflection amplitude for
graphene (substrate)  is determined by \cite{Volokitin2001b}
\begin{equation}
R_{g(d)}=\frac{\epsilon _{g(d)}-1}{\epsilon _{g(d)}+1},
 \label{refcoef}
\end{equation}
where $\epsilon _{g(d)}$ is the dielectric function for graphene
(substrate).

In the study below we used the dielectric function of graphene, which was
calculated recently within the random-phase approximation (RPA)
\cite{Wunsch2006,Hwang2007}. The small (and constant) value of
the graphene Wigner-Seitz radius $r_s$ indicates that it is a weakly
interacting system for all carries densities, making the RPA an
excellent approximation for graphene (RPA is asymptotically
exact in the $r_s\ll1$ limit).  The dielectric function
is an analytical function in the upper half-space
of the complex $\omega$-plane:
\begin{equation}
\epsilon_g(\omega,q)=1+\frac{4k_Fe^2}{\hbar
v_Fq}-\frac{e^2q}{2\hbar \sqrt{\omega^2-v_F^2q^2}}\Bigg \{G\Bigg
(\frac{\omega+2v_Fk_F}{v_Fq}\Bigg )- G\Bigg
(\frac{\omega-2v_Fk_F}{v_Fq}\Bigg )-i\pi \Bigg \},
\end{equation}
where
\begin{equation}
G(x)=x\sqrt{1-x^2} - \ln(x+\sqrt{1-x^2}),
\end{equation}
where the Fermi wave vector $k_F=(\pi n)^{1/2}$, $n$ is the
concentration of charge carriers, the Fermi energy
$\epsilon_F=\gamma k_F=\hbar v_Fk_F$, $\gamma=\hbar v_F\approx
6.5$ eV\AA, and $v_F$ is the Fermi velocity.

The dielectric function of amorphous SiO$_2$ can be described
using an oscillator model\cite{Chen2007}
\begin{equation}
\epsilon(\omega) =
\epsilon_{\infty}+\sum_{j=1}^2\frac{\sigma_j}{\omega_{0,j}^2-\omega^2-i\omega\gamma_j},
\end{equation}
where parameters $\omega_{0,j}$, $\gamma_j$ and $\sigma_j$ were
obtained by fitting the actual $\epsilon$ for SiO$_2$ to the above
equation, and are given by $\epsilon_{\infty}=2.0014$,
$\sigma_1=4.4767\times10^{27}$s$^{-2}$, $\omega_{0,1}=8.6732\times
10^{13}$s$^{-1}$, $\gamma_1=3.3026\times 10^{12}$s$^{-1}$,
$\sigma_2=2.3584\times10^{28}$s$^{-2}$, $\omega_{0,2}=2.0219\times
10^{14}$s$^{-1}$, and $\gamma_1=8.3983\times 10^{12}$s$^{-1}$.

The equilibrium or steady state temperature can be obtained from the condition that
the heat power generated by friction must be equal to the heat transfer
across the substrate surface
\begin{equation}
F_x(T_d,T_g)v=S_z(T_d,T_g)+\alpha_{ph}(T_g-T_d), \label{eqtemp}
\end{equation}
where the second term in Eq. (\ref{eqtemp}) takes into account the
heat transfer through direct phononic coupling; $\alpha_{ph}$
is the thermal contact conductance due to phononic coupling.

\textbf{Results.} At small velocities ($v\ll v_F$) the friction
force depends linearly on $v$: $F_x=\Gamma v$, where the friction
coefficient $\Gamma$ is determined by
\begin{equation}
\Gamma =\frac \hbar {2\pi ^2} \int_0^\infty d\omega
\Bigg(-\frac{\partial n_d(\omega)}{\partial \omega} \Bigg )\int_{0
}^\infty dq q^3e^{-2qd}
\frac{\mathrm{Im}R_{d}(\omega)\mathrm{Im}R_{g}(\omega) }{\mid
1-e^{-2 q d}R_{d}(\omega)R_{g}(\omega)\mid ^2}\label{frcoef}
\end{equation}
 The low-field mobility can be described using
Matthiessen's rule
\begin{equation}
\mu^{-1} = \mu_{int}^{-1} + \mu_{ext}^{-1},
\end{equation}
where $\mu_{int}$ is the intrinsic mobility due to scattering of
the graphene free carriers against the acoustic and optical
phonons in graphene, and $\mu_{ext}$ is the extrinsic mobility due
to scattering against the optical phonons in the substrate.
According to the theory of the van der Waals friction $\mu_{ext} =
ne/\Gamma$. Scattering of the graphene carriers by the acoustic
phonons of graphene places an intrinsic limits on the low-field
room temperature mobility given by $\mu=200000$cm$^2$/Vs at the
carriers density 10$^{12}$cm$^{-2}$    \,\cite{Chen2008}. However,
for graphene on SiO$_2$ the mobility of the graphene carriers is
reduced to $\mu=40000$cm$^2$/Vs. This reduction must result partly
from interface imperfections, and partly from scattering of the
graphene carriers from the optical phonons of the SiO$_2$
substrate.

\begin{figure}
\includegraphics[width=0.70\textwidth]{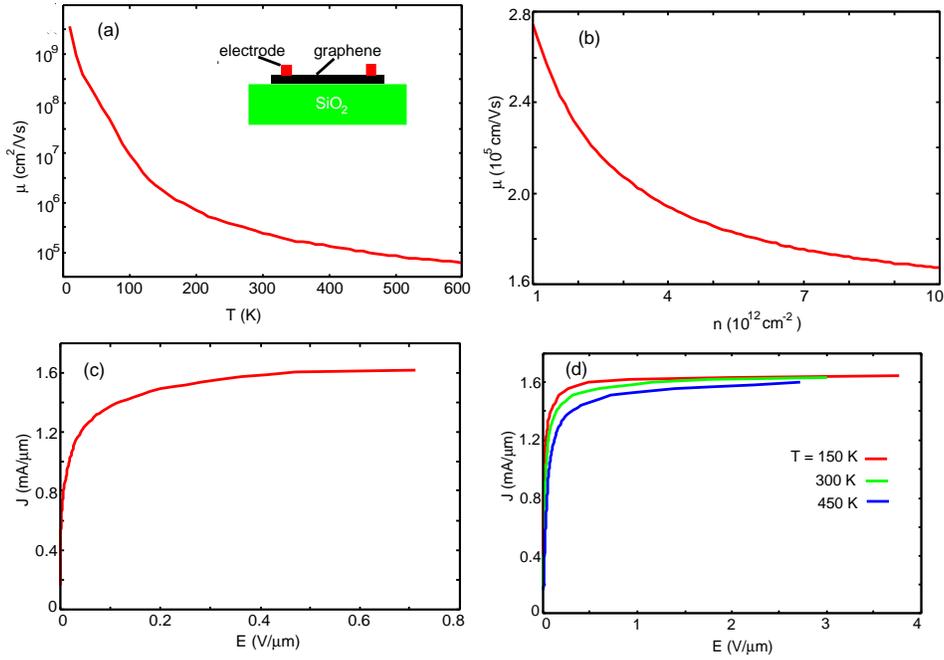}
\caption{\label{Fig2} The role of the interaction between phonon
polaritons in SiO$_2$ and free carriers in graphene for graphene
field-effect transistor transport. The separation between graphene
and SiO$_2$ is $d=3.5$\AA. (a) Dependence of the low-field
mobility on the temperature. Concentration of charge carriers
$n=10^{12}$cm$^{-12}$. (b) Dependence  of the low-field mobility
on the concentration of charge carriers $n$ for $T_d=T_g=300$ K.
(c) Current density-electric field dependence at $T=0$ K,
$n=10^{12}$cm$^{-12}$. (d) The same as in (c) but for different
temperatures.}
\end{figure}

Fig. \ref{Fig2}a shows the calculated dependence of the low-field
mobility on the temperature, due to scattering from the optical
phonons in SiO$_2$. The calculations were performed using the
theory of the van der Waals friction. The mobility is
characterized by a strong temperature dependence. At room
temperature the calculated mobility $\mu \approx 3\times 10^5 \
{\rm cm^2 /Vs}$, which is approximately one order of magnitude
larger than the mobility calculated in Ref. \cite{Perebeinos2010}.
This indicates that for graphene adsorbed on SiO$_2$ the main
contribution to the low-field mobility comes not from scattering
from the optical phonons in SiO$_2$, but from scattering from
point defects on the interface.

Fig. \ref{Fig2}b shows the dependence of the mobility at $T_d=300$
K on the charge density $n$. The mobility
monotonically decreases when the concentration of charge carriers
increases. Figs. \ref{Fig2}c and \ref{Fig2}d show the dependence
of the current density on the electric field at
$n=10^{12}$cm$^{-2}$, and for different temperatures.  In
obtaining these curves we have used that $J=nev$ and
$neE=F_x$, where $J$ and $E$ are current density and electric
field, respectively.  Note that the current density saturate at
$E\sim 0.5-2.0$V/$\ \mu$m, which is in agreement  with
experiment \cite{Freitag2009}. The saturation velocity
can be extracted from the $I-E$ characteristics using
$J_{sat}=nev_{sat}$, where 1.6 mA/$\mu$m is the saturated current
density, and with the charge density concentration
$n=10^{12}$cm$^{-2}$: $v_{sat}\approx 10^6$m/s. Fig. \ref{Fig2}c
was calculated at $T_d=0$ K. At zero temperature the van der Waals
friction is due to quantum fluctuations of charge density, and is
determined by the second term in Eq. (\ref{parallel2})
\cite{Pendry1997,Volokitin99,VolokitinRMP07,VolokitinUFN07,Volokitin2008b}
\begin{equation}
F_x(T_d=T_g=0) =-\frac\hbar {\pi ^3}\int_{0 }^\infty
dq_y\int_0^\infty dq_x\int_0^{q_xv}d\omega q_xe^{-2qd}
\frac{\mathrm{Im}R_{d}(\omega)\mathrm{Im} R_{g}(\omega^-)} {\mid
1-e^{-2qd}R_{d}(\omega)R_{g}(\omega^-)\mid ^2}
\end{equation}
The existence of ``quantum'' friction is still debated in the
literature \cite{Philbin2009,Pendry2010,Volokitin2010}. The theory
\cite{Pendry1997,Volokitin99,VolokitinRMP07,VolokitinUFN07,Volokitin2008b}
predicts that a solid moving relative to another
experiences a force due to quantum fluctuations, that is
opposite to the direction of motion.
The van der Waals friction can be studied in non-contact
experiments, and in frictional drag experiments
\cite{VolokitinRMP07,VolokitinUFN07}. In both these experiments
the solids are separated by a potential barrier thick enough to
prevent electrons or other particle with a finite rest mass from
tunneling across it, but allowing the interaction via the
long-range electromagnetic field, which is always present in the
gap between bodies. In non-contact friction experiments the
damping of cantilever vibrations is typically measured, while in
frictional drag experiments a current density is induced in one
medium. The friction between the moving charge carriers and
nearby medium gives rise to a change of $I-E$ characteristics,
which can be measured.

The friction force acting on the charge carriers in graphene for
high electric field is determined by the interaction with the
optical phonons of the graphene, and with the optical phonons of
the substrate. The frequency of optical phonons in graphene are a
factor 4 larger than for the optical phonon in SiO$_2$. Thus, one
can expect that for graphene on SiO$_2$ the high-field $I-E$
characteristics will be determined by excitations of optical
phonons in SiO$_2$. According to the theory of the van der Waals
friction\cite{VolokitinRMP07,VolokitinUFN07}, the ``quantum''
friction, which exist even at zero temperature, is determined by
the creation of excitations in each of the interacting medium, the
frequencies of which are connected by $vq_x=\omega_1 + \omega_2$.
The relevant excitations in graphene are the electron-hole pairs
which frequencies begining from zero, while for SiO$_2$ the
frequency of surface phonon polaritons $\omega_{ph} \approx
60$meV. The characteristic wave vector of graphene is determined
by Fermi wave vector $k_F$.  Thus the friction force is strongly
enhanced when $v>v_{sat}=\omega_{ph}/k_F \sim 10^6$m/s, in the
accordance with numerical calculations. Thus the measurements of
the current density-electric field relation of graphene adsorbed
on SiO$_2$ give the possibility to detect ``quantum'' friction,
what has fundamental significance for physics.

\begin{figure}
\includegraphics[width=0.45\textwidth]{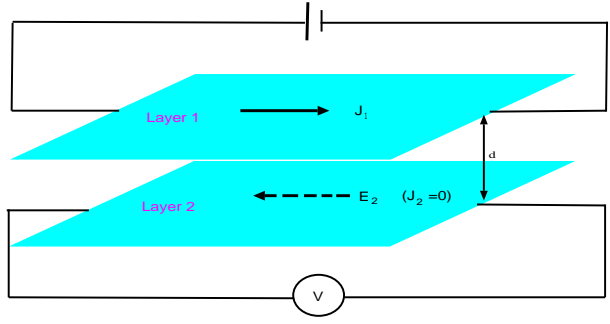}
\caption{\label{Fig3} A schematic diagram of a drag experiment. A
current $\mathbf{J_1}$ is passed through layer \textbf{1}, and a
voltmeter is attached to layer \textbf{2} to measure the induced
electric field $\mathbf{E_2}$ due to the inter-layer interaction.}
\end{figure}

An alternative method of studying of the van der Waals friction
consists in driving an electric current in one metallic layer and
studying of the effect of the frictional drag on the electrons in
a second (parallel) metallic layer (Fig. \ref{Fig3}).
\begin{figure}
\includegraphics[width=0.70\textwidth]{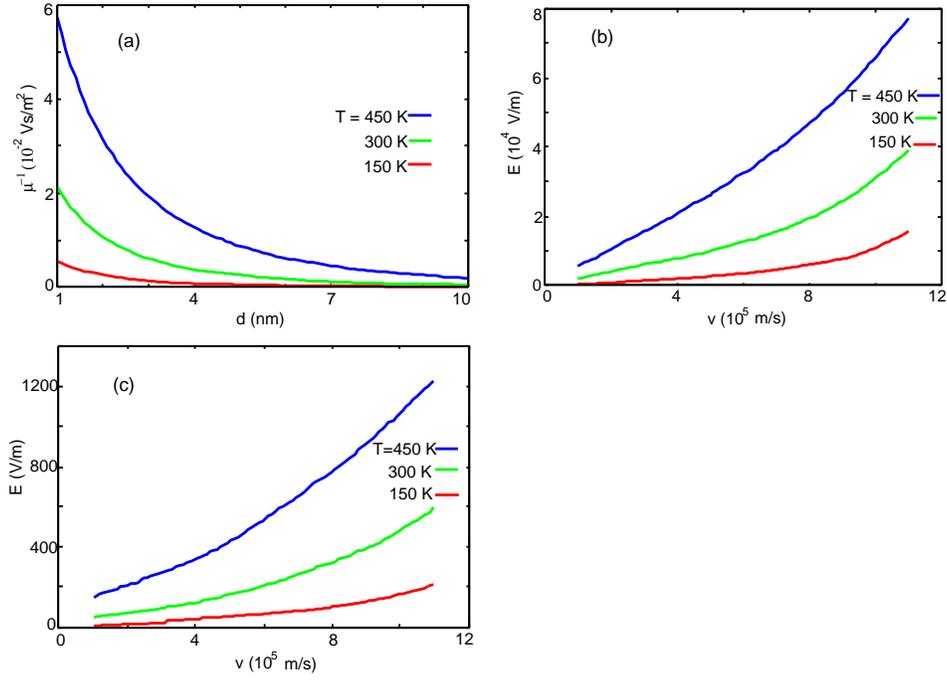}
\caption{\label{Fig4} Frictional drag between two graphene sheets
at the carrier concentration $n=10^{12}$cm$^{-2}$.  (a)
Dependence of friction coefficient per unit charge,
$\mu^{-1}=\Gamma/ne$, on the separation between graphene sheets $d$.
(b) Dependence of electric field induced in graphene on drift
velocity of charge carriers in other graphene sheet at
the layer separation $d=1$ nm. (c) The same as in (b) but at
$d=10$nm.}
\end{figure}
Such experiments were first suggested by Pogrebinskii
\cite{Pogrebinskii} and Price \cite{Price}, and were performed for
2D-quantum wells \cite{Gramila1,Sivan}. In these experiments, two
quantum wells are separated by a dielectric layer thick enough to
prevent electrons from tunneling across it, but allowing
inter-layer interaction between them. A current  density $J_1=n_1
ev$ is driven through layer \textbf{1} (where $ n_1 $ is the
carrier concentration per unit area in the first layer), see
 Fig. \ref{Fig3}. Due to  the
inter-layer interactions a frictional stress $\sigma =\Gamma v$
will act on the electrons in layer \textbf{2} from layer
\textbf{1}, which will induce a current in layer \textbf{2}. If
layer \textbf{2} is an open circuit, an electric field $E_2$ will
develop in the layer whose influence cancels the frictional stress
$\sigma$ between the layers. Experiments \cite{Gramila1} show
that, at least for small separations, the frictional drag can be
explained by the interaction between the electrons in the
different layers via the fluctuating Coulomb field. However, for
large inter-layer separation the frictional drag is dominated by
phonon exchange \cite{Bonsager}.

Similar to 2D-quantum wells in semiconductors, frictional drag
experiments can be performed (even more easy) between graphene
sheets. Such experiments can be performed in vacuum where the
contribution from phonon exchange can be excluded. To exclude
noise (due to presence of dielectric) the frictional drag
experiments between quantum wells were performed at very low
temperature ($T\approx 3 \ {\rm K}$)\cite{Gramila1}. For suspended
graphene sheets there are no such problem and frictional drag
experiment can be performed at room temperature. In addition,
2D-quantum wells in semiconductors have very low Fermi energy
$\epsilon_F \approx 4.8\times 10^{-3}$eV \cite{Gramila1}. Thus
electrons in these quantum wells are degenerate only for very low
temperatures $T<T_F=57$ K. For graphene the Fermi energy
$\epsilon_F=0.11$eV at $n=10^{12}$cm$^{-2}$, and the electron gas
remains degenerate for $T<1335$ K.

At small velocities the electric field induced by frictional drag
depends linear on the velocity, $E=(\Gamma/ne)v = \mu^{-1}v$,
where $\mu$ is the low-field mobility. For $\hbar \omega \ll
\epsilon_F$ and $q\ll k_F$ the reflection amplitude for graphene
is given by the same expression as for a 2D-quantum well
\cite{VolokitinRMP07}:
\begin{equation}
R_g                                 =1+\frac{\hbar
\omega}{2k_Fe^2}\label{refcoef}
\end{equation}
Substitution of Eq. (\ref{refcoef}) into Eq. (\ref{frcoef}) gives
\begin{equation}
\Gamma =
0.01878\frac{\hbar}{d^4}\left(\frac{k_BT}{k_Fe^2}\right)^2
\end{equation}
\begin{figure}
\includegraphics[width=0.70\textwidth]{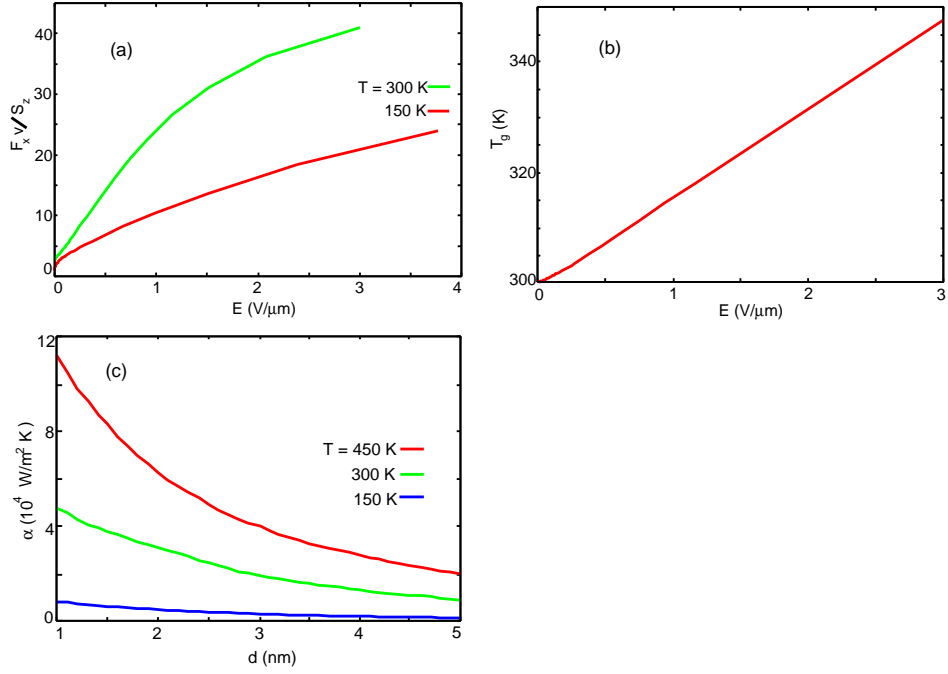}
\caption{\label{Fig5} Radiative heat transfer between graphene and
SiO$_2$. (a) The dependence of the ratio between the total heat
generated by current and the radiative heat flux, on electric
field. $n=10^{12}$cm$^{-2}$, $d=3.5$\AA, $\alpha_{ph}=1.0\times
10^8$Wm$^{-2}$K$^{-1}$ (b) Dependence of graphene temperature
$T_g$ on the electric field. The substrate temperature $T_d=300$
K. Parameters are the same as in (a). (c) Dependence of the
thermal contact conductance on the separation $d$ for low electric
field, $\alpha_{ph}=0$.}
\end{figure}
Fig. \ref{Fig4}a shows the dependence of the friction coefficient (per
unit charge) $\mu^{-1}$ on the separation $d$ between the sheets. For
example, $E=5\times 10^{-4}v$ for $T=300$ K and $d=10$ nm. For
a graphene sheet of length $1\ {\rm \mu m}$, and with $v=100$m/s this
electric field will induce the voltage $V=10$ nV. Figs. \ref{Fig5}b  and \ref{Fig5}c
shows the induced electric field-velocity relation for high velocity, with
$d=1$nm (b) and $d=10$nm (c).

As discussed above, for graphene on SiO$_2$ the
excess heat generated by the current is transferred to the substrate
through the near-field radiative heat transfer,
and via the direct phononic coupling (for which the thermal
contact conductance $\alpha \approx 10^8$Wm$^{-2}$K$^{-1}$).  At
small temperature difference ($\Delta T=T_g - T_d\ll T_d$), from
Eq. (\ref{eqtemp}) we get
\begin{equation}
\Delta T=
\frac{F_{x0}v-S_{z0}}{\alpha_{ph}+S_{z0}^{\prime}-F_{x0}^{\prime}v}
\label{eqtemp1}
\end{equation}
where $F_{x0}=F_x(T_d,T_g=T_d)$, $S_{z0}=S_z(T_d,T_g=T_d)$,
\[
F_{x0}^{\prime}=\frac{dF_x(T_d,T_g)}{dT_g}\Big|_{T_g=T_d}, \,\,
S_{z0}^{\prime}=\frac{dS_z(T_d,T_g)}{dT_g}\Big|_{T_g=T_d}
\]
The thermal contact conductance is given by
\begin{equation}
\alpha =\frac {S_z(T_d,T_g)+\alpha_{ph}\Delta T}{\Delta T}
\approx
\frac{(\alpha_{ph}+S_{z0}^{\prime})F_{x0}v-S_{z0}F_{x0}^{\prime}v}{F_{x0}v-S_{z0}}
\label{alpha}
\end{equation}
From Eq. (\ref{alpha}) it follows that the thermal contact
conductance can be strongly enhanced when $F_{x0}v\approx S_{z0}$.
Fig. \ref{Fig5}a shows the ratio of the total heat generated by
the current to the radiative heat flux. For low field this ratio
$\sim$ 2 so that in this case both radiative heat transfer and
phononic heat transfer give contributions of the same order. The
phononic mechanism of heat transfer dominates for high electric
field. For high electric field the equilibrium graphene
temperature due to self-heating depends linear on the electric
field (see Fig. \ref{Fig5}b) in contrast to the low-field case
where this dependence is quadratic. This can be explains by the
fact that the saturation velocity depends weakly on electric
field; therefore heat, generated by friction, will linearly depend
on the electric field. Fig. \ref{Fig5}c shows the dependence of
the  thermal contact conductance on the separation $d$ for low
electric field. At $d \sim$ 5 nm and $T=300$ K the thermal contact
conductance, due to the near-field radiative heat transfer, is
$\sim 10^4$W$\cdot$m$^{-2}\cdot$K$^{-1}$, which is $\sim 3$ orders
of magnitude larger than that of the black-body radiation. In
comparison, the near-field radiative contact conductance in
SiO$_{2}$-SiO$_{2}$ for the plate-plate configuration, when
extracted from experimental data \cite{Chen2009a} for the
plate-sphere configuration, is $\sim $
2230W$\cdot$m$^{-2}\cdot$K$^{-1}$ at a $\sim$30 nm gap. For this
system the thermal contact conductance depends on separation as
$1/d^2$; thus $\alpha \sim 10^5$W$\cdot$m$^{-2}\cdot$K$^{-1}$ at
$d\sim 5$ nm what is one order of magnitude larger than for the
graphene-SiO$_2$ system in the same configuration. However, the
sphere has a characteristic roughness of $\sim$ 40 nm, and the
experiments \cite{Chen2009a,Greffet2009} were restricted to
separation wider than 30 nm (at smaller separation the
imperfections affect the measured heat transfer). Thus the extreme
near-field-separation, with $d$ less than approximately 10 nm, may
not be accessible using a plate-sphere geometry. On the other hand
suspended graphene sheet has a roughness $\sim$1 nm
\cite{Meyer2007}, and measurements of the thermal conductance can
be performed from separation larger than $\sim$ 1 nm. At such
separation one would expect the emergency of nonlocal and
nonlinear effects. This range is of great interest for the design
of nanoscale devices, as modern nanostructures are considerably
smaller than 10 nm and are separated in some cases by only a few
Angstroms. Another advantage of using graphene for studying the
radiative heat transfer result from the fact that under
steady-state condition the heat flow is equal to the heat
generated in the graphene by the current density: $S_z=EJ$. This
quantity can be accurately obtained from $I-V$ characteristics.
The temperature of graphene can be measured accurately using Raman
scattering spectroscopy \cite{Meyer2007}.

\textbf{Conclusion.} We have used theories of the near-field
radiative heat transfer and the van der Waals friction to study
transport, heat generation and dissipation in graphene due to
the interaction with phonon-polaritons in the (amorphous) SiO$_2$
substrate. In contrast with existing theories, based on the
semiclassical Boltzmann transport
equation, our approach is macroscopic. In the latter approach all
microscopic properties are included in dielectric functions of
material. Explicit formulas were obtained for the low-field mobility,
the high-field transport and the near-field radiative heat transfer
between graphene and the substrate. The low-field mobility exhibit a strong
temperature dependence, which (according to the theory of
the van der Waals friction) is associated with the
thermal fluctuation inside the media.
High-field transport exhibit a weak temperature
dependence, which can be considered as manifestation of quantum
fluctuations. Thus the study of transport properties in graphene
gives the possibility to detect ``quantum'' friction, the existence of
which is still debated in literature. We have calculated the
frictional drag between graphene sheets mediated by the van der Waals
friction, and found that it can induce large enough voltage
to be easily measured experimentally. This effect can be used in
electronics and for sensoring. We have shown that for the low-field
heat transfer between graphene and the substrate, both radiative heat
transfer and phononic heat transfer give contributions of the same
order. High-field heat transfer is determined by the phononic
mechanism. We have pointed out that graphene can be used to study
near-field radiative heat transfer in the plate-plate configuration,
and for shorter separations than it is possible now in the
plate-sphere configuration.

 \vskip 0.5cm
\textbf{Acknowledgment}

A.I.V acknowledges financial support from the Russian Foundation
for Basic Research (Grant N 08-02-00141-a) and ESF within activity
``New Trends and Applications of the Casimir Effect''.


\vskip 0.5cm






\begin{thebibliography}{999}


\bibitem{Geim2004} Novoselov, K.S.; Geim, A.K.; Morosov, S.V.; Jiang, D.; Zhang, Y.; Dubonos, S.V.;
Grigorieva, I.V.; Firsov, A.A. \textit{Science} \textbf{2004}, \textit{306}, 666-669.

\bibitem{Geim2005} Novoselov, K.S.; Geim, A.K.; Morosov, S.V.; Katsnelson, M.I.; Grigorieva, I.V.; Dubonos, S.V.;
Firsov, A.A. \textit{Nature (London)} \textbf{2005}, \textit{197},
197-200.

\bibitem{Geim2007} Geim, A.K.; Novoselov, K.S. \textit{Nat. Mater.} \textbf{2007}, \textit{6}, 183-191.

\bibitem{Rytov53}   Rytov, S. M. \textit{Theory of Electrical
Fluctuation and Thermal Radiation} (Academy of Science of USSR
Publishing, Moscow, 1953)

\bibitem{Rytov67}   Levin, M. L.;  Rytov, S. M. \textit{Theory of eqilibrium thermal
fluctuations in electrodynamics} (Science Publishing, Moscow,
1967)

\bibitem{Rytov89}   Rytov, S. M.;  Kravtsov, Yu. A.;  Tatarskii, V. I. \textit{
Principles of Statistical Radiophyics}(Springer, New York.1989),
Vol.3




\bibitem{Lifshitz54}  E. M. Lifshitz \textit{Zh. Eksp. Teor. Fiz.}  \textbf{1955},\textit{29 },
94-110 [Sov. Phys.-\textit{JETP} \textbf{1956}, \textit{2}, 73-83]

\bibitem{Polder1971}  Polder,  D.;  Van Hove, M. \textit{Phys. Rev. B} \textbf{1971}, \textit{4},
3303-3314



\bibitem{Volokitin99}   Volokitin, A.I.; Persson,B.N.J. \textit{J.Phys.:
Condens. Matter} \textbf{11}, 345
(1999);\textit{Phys.Low-Dim.Struct.}\textbf{7/8},17 (1998)

\bibitem{VolokitinRMP07}  Volokitin, A. I.;  Persson, B. N. J. \textit{Rev. Mod. Phys.} \textbf{2007}, \textit{79},
1291-1329

\bibitem{VolokitinUFN07} Volokitin, A. I.;  Persson, B. N. J. \textit{Usp. Fiz.
Nauk} \textbf{2007}, \textit{177}, 921-951 [\textit{Phys. Usp.}
\textbf{2007}, \textbf{50}, 879-906


\bibitem{Volokitin2008b} Volokitin, A.I.; Persson, B.N.J. \textit{Phys. Rev. B} \textbf{2008},
\textit{
78}, 155437.



\bibitem{Volokitin2001b}  Volokitin, A.I.; Persson, B.N.J. \textit{J.Phys.:
Condens. Matter} \textbf{2001}, \textit{13}, 859-873

\bibitem{Volokitin2008a} Volokitin, A.I.; Persson, B.N.J. \textit{Phys. Rev. B} \textbf{2008},\textit{77}, 033413



\bibitem{Perebeinos2009} Rotkin, S.V.; Perebeinos, V.; Petrov,
A.G.; Avouris, P. \textit{Nano Lett.} \textbf{2009}, \textit{9},
1850-1855

\bibitem{Perebeinos2010} Perebeinos, V.; Avouris, P. \textit{Phys.
Rev. B} \textbf{2010}, \textit{81}, 195442

\bibitem{Handbook2007}  \textit{Handbook of Nanoscience, Engineering and Technology}; Goddard, W., Brenner, D.,
Lishevski. S., Iafrate. G.J., Eds.; Taylor and Francis-CRC Press:
Boca Raton, FL, 2007


\bibitem{Persson2010} Persson, B.N.J.; Ueba, H. \textit{Europhys. Lett.}
\textbf{2010}, 91, 56001.

\bibitem{Chen2009} Chen, D.Z.A.; Haman, R.; Soljacic, M.; Joannopoulos, J.D.; Chen, G. \textit{Appl. Phys. Lett.}
\textbf{2007}, 90, 181921

\bibitem{Chen2009a} Shen, S.; Narayanaswamy, A.; Chen, G. \textit{Nano Lett.} \textbf{2009}, 9, 1883-1888.

\bibitem{Greffet2009} Rousseau, E.; Siria, A.; Jourdan, G.; Volz, S.; Comin, F.; Chevrier, J.; Greffet, J.J.
\textit{Nature Photon.} \textbf{2009}, \textit{90}, 181921

\bibitem{Greffet6}   Joulain, K.;Mulet, J.P.;Marquier, F.;Carminati, R.;
Greffet, J.J. \textit{Surf. Sci. Rep.} \textbf{2005} ,
\textit{57,}59-112



\bibitem{Wunsch2006} Wunscvh, B.; Stauber, T.; Sols, F.; Guinea, F.; \textit{New J.Phys.} \textbf{2006},
\textit{8}, 318.


\bibitem{Hwang2007} Hwang, E.H.; Sarma, S.D.; \textit{Phys. Rev. B} \textbf{2007}, \textit{75}, 205418

\bibitem{Chen2007} Chen, D.Z.A.; Hamam, R,; Solia\u ci\,c,
M.;Joannopoulos, J.D. \textit{Appl. Phys. Lett.} \textbf{2007},
\textit{90}, 181921

\bibitem{Chen2008}  Chen, J.H.; Jang, C.; Xiao, S.; Ishigami, M.;
Fuhrer, M.S. \textit{Nat. Nanotechnol.} \textbf{2008}, \textit{3},
206

\bibitem{Freitag2009} Freitag, M.; Steiner, M.; Martin, Y.;
Perebeinos, V.; Chen, Z.; Tsang, J.C.; Avouris, P. \textit{Nano
lett.}, \textbf{2009}, \textit{9}, 1883-1888

\bibitem{Pendry1997} Pendry, J.B. \textit{J. Phys.C},  \textbf{1997}, \textit{9},
10301-10320

\bibitem{Philbin2009} Philbin, T.G.; Leonhardt, U. \textit{New J.
Phys.} \textbf{2009}, \textit{11}, 033035

\bibitem{Pendry2010} Pendry, J.B. \textit{New J.
Phys.} \textbf{2010}, \textit{12}, 033028

\bibitem{Volokitin2010}  Volokitin, A.I.; Persson, B.N.J. to be
published



\bibitem{Pogrebinskii}   Pogrebinskii, M.B. \textit{Fiz.Tekh.Poluprov.} \textbf{1977},\textit{11},
637-641 [\textit{Sov.Phys. Semicond.} \textbf{1977}, \textit{11},
372-376].

\bibitem{Price}   Price, P. J. \textit{Physica B+C},  \textbf{1983}, \textit{117},
750-752


\bibitem{Gramila1}   Gramila, T.J.;  Eisenstein, J.P.;  MacDonald, A.H.;
Pfeiffer, L.N.; West, K. W. \textit{Phys. Rev. Lett.}
\textbf{1991}, \textit{66}, 1216-1219

\bibitem{Sivan}   Sivan, U.;  Solomon, P.M.;  Shtrikman, H. \textit{Phys. Rev.
Lett.} \textbf{1992}, \textit{68, }1196-1199

\bibitem{Bonsager}  B{\o}nsager, M.C.;  Flensberg, K.;  Hu, B. Y.K.;  Macdonald, A.H. \textit{Phys.Rev.
B}, \textbf{1998}, \textit{57}, 7085-7102

\bibitem{Meyer2007} Meyer, J.C.; Geim, A.K.; Katsnelson, M.I.;
Novoselov, K.S.; Booth, T.J.; Roth, S. \textit{Nature},
\textbf{2007}, \textit{446}, 60-63























\end{thebibliography}
\end{document}